\documentclass[12pt,fleqn]{article}
\usepackage{verbatim,amssymb,amsmath,graphics,cite}

\newcommand{\ci}{\cite}

\newcommand{\HS}{\ensuremath{\mathcal{H}}}
\newcommand{\Phx}{\ensuremath{\Phi^\times}}
\newcommand{\bk}[2]{\ensuremath{\langle #1|#2 \rangle}}
\newcommand{\kt}[1]{\ensuremath{|#1\rangle}}

\newcommand{\pbk}[2]{\ensuremath{\langle {}^{+}#1|#2{}^{+} \rangle}}
\newcommand{\pkt}[1]{\ensuremath{|#1{}^{+}\rangle}}

\newcommand{\mbk}[2]{\ensuremath{\langle {}^{-}#1|#2^{-} \rangle}}
\newcommand{\mkt}[1]{\ensuremath{|#1{}^{-}\rangle}}

\newcommand{\pmkt}[1]{\ensuremath{|#1{}^{\pm}\rangle}}

\begin{document}

\begin{titlepage}
\begin{center}

\huge \textbf{On the Mass and Width of the $Z$-boson and Other
Relativistic Quasistable Particles} \vspace{1cm}\\ \Large
\textbf{Arno~R.~Bohm\footnote{E-mail: bohm@physics.utexas.edu} and
N.~L.~Harshman\footnote{E-mail: harshman@physics.utexas.edu}} 
\vspace{1cm}\\ \small{The University of Texas at Austin}\\
\small{Austin, Texas~78712}

\begin{abstract}

The ambiguity in the definition for the mass and width of relativistic
resonances is discussed, in particular for the case of the $Z$-boson.
This ambiguity can be removed by requiring that a resonance's width
$\Gamma$ (defined by a Breit-Wigner lineshape) and lifetime $\tau$
(defined by the exponential law) always and exactly fulfill the
relation $\Gamma = \hbar/\tau$.  To justify this one needs
relativistic Gamow vectors which in turn define the resonance's mass
$M_R$ as the real part of the square root $\rm{Re}\sqrt{s_R}$ of the
$S$-matrix pole position $s_R$.  For the $Z$-boson this means that $M_R
\approx M_Z - 26\mbox{MeV}$ and $\Gamma_R \approx
\Gamma_Z-1.2\mbox{MeV}$ where $M_Z$ and $\Gamma_Z$ are the values
reported in the particle data tables.

\end{abstract}
\end{center}



\normalsize

\end{titlepage}

\section{Introduction}

Relativistic resonances and quasi-stable particles are listed along
with stable, elementary particles \ci{PDG} and are not considered
qualitatively different from the latter.  Both are characterized by
species labels or internal quantum numbers, spin-parity $j^P$ and the
center of mass energy squared $s$.  The only difference between the
characterizations is that stable particles have a real value for
$s=m^2 \geq 0$ and quasi-stable particles and resonances have a
complex value for $s$, which is often also parameterized by the two
real values ``mass'' 
$m$ and ``width'' $\Gamma$.  These can be combined  as
$s=(m-i\frac{\Gamma}{2})^2$ among other ways\footnote{We use the
notation $m$, $\Gamma$ if we do not specify which mass and width
parameter is meant.  $M_R$, $\Gamma_R$ is used for the complex pole
position $s_R=(M_R-i\frac{\Gamma_R}{2})^2$ of the $S$-matrix or the pole
position of the $Z$ propagator.  Conventionally one parameterizes the pole
position using the quantities $M_{\rho}$, $\Gamma_{\rho}$ given by 
\begin{equation}\label{rhoR}
s_R \equiv M_{\rho}^2 - iM_{\rho}\Gamma_{\rho} =
M_R^2(1-\frac{1}{4}(\frac{\Gamma_R}{M_R})^2) - iM_R \Gamma_R 
\end{equation}
and calls $M_{\rho} = M_R
\sqrt{1-\frac{1}{4}(\frac{\Gamma_R}{M_R})^2}$ the resonance mass and
$\Gamma_{\rho} = \Gamma_R
(1-\frac{1}{4}(\frac{\Gamma_R}{M_R})^2)^{-\frac{1}{2}}$ its width.
To denote $m$ and  $\Gamma$ in the on-mass-shell definition we use
$M_Z$ and $\Gamma_Z$.  The notation in the literature varies (some call
$M_Z \rightarrow M_R$ or vice versa).}.

The ratio of the resonance characterization parameter $(\Gamma/m)$ runs
over a wide range: from $(\Gamma/m) \sim 10^{-2}$-$10^{-1}$ for hadron
resonances
and $(\Gamma/m) \sim 3 \times 10^{-2}$ for the $Z$-boson to much smaller
values for other electroweakly decaying particles,
e.g. $(\Gamma/m)_{\pi^0} \sim 10^{-7}$, $(\Gamma/m)_{\pi^\pm} \sim
10^{-15}$ and $(\Gamma/m)_{K^0} \sim 10^{-14}$.
The experimental definition or method of determination
of $\Gamma$ and $m$ is quite different for different magnitudes of
$(\Gamma/m)$.

\begin{enumerate}

\item For $\Gamma/m>10^{-7}$, $\Gamma$ and $m$ are determined as the
lineshape parameters of relativistic Breit-Wigner (BW) amplitudes from the
cross sections and asymmetries (e.g. for the $Z$-boson in $e\bar e\to Z\to
f\bar f$ \ci{Riemann}).  In what follows we discuss two different
relativistic BW's.

\item For $\Gamma/m\sim10^{-14}$ and less, the lineshape cannot be
resolved and instead of $\Gamma$ one measures the lifetime $\tau$.
The definition of $\tau$ is based on the exponential law
for the partial decay rates $\dot P_{\eta}(t)$ and/or the decay
probability $P(t)=\sum_{\eta} P_\eta(t)$. One measures $\tau$ by a fit to
the exponential law~\cite{exponential.fit}. 
\end{enumerate}

Observationally, $\Gamma$ and $\tau$ are different qualities; $\Gamma$
comes from the BW energy distribution and $\tau$ from the exponential
counting rate.  Statements that the exponential law is not exact and the
use of alternate forms for the relativistic BW have further
confused the connection between the width and the lifetime.
Connected with the definition of resonance width is the definition of
resonance mass. Experimentalists have in the past identified it with
the peak of the invariant mass distribution which was sufficient since
the data where not accurate enough.  $S$-matrix theoreticians prefer
to define it as $\rm{Re}\sqrt{s_R}=M_R$ instead of the peak
$\sqrt{\rm{Re}\,s_R}=M_\rho$ or the on-shell definition $M_Z$ (see
footnote 1).  Only since the LEP data for the $Z$-boson has accuracy
reached a level where these differences have become of practical
importance.  Nevertheless, the exact definition
of $m$ and $\Gamma$ and the relation of $\Gamma$ to $\tau$ have always
been of fundamental importance.  Conventionally, the identification 
$\hbar/\Gamma \approx \tau$ is 
justified by some heuristic and approximate
arguments~\cite{Goldberger.Watson,review} based on the  
Weisskopf-Wigner (WW) methods \cite{Weisskopf.Wigner}. So
far, no generally 
accepted theory precisely related $\hbar/\Gamma$ and $\tau$
because there did ``not exist $\ldots$ a rigorous theory to which
these various methods [WW, etc.] can be considered as approximations''
\ci{Levy}.

A rigorous description of non-relativistic resonances and decaying
states by Gamow kets $\psi^G=\mkt{E-i\Gamma/2}$ has been
provided during the last two decades by the Rigged Hilbert Space (RHS)
formulation of quantum mechanics \ci{Bohm.etal,Harshman, Bohm2}.
For these Gamow kets one can derive both an exact Breit-Wigner with width
$\Gamma$ for the lineshape and an exact, exponential time evolution
$\psi^G(t)=e^{-iEt}e^{-\frac{\Gamma}{2}t}\psi^G(0)$.  From the latter follows
the exact Golden Rule for the partial
decay rate of the resonance $R$ into the channel $\eta$, $\dot{\cal
P}_{\eta}(t)=\Gamma_{\eta}e^{-\Gamma t}$. Thus with the properties of
the Gamow vectors, the precise relation $\tau =
\hbar/\Gamma$ between the differently defined quantities $\tau$ and
$\hbar/\Gamma$ was established for non-relativistic resonances.

Guided and motivated by recent discussions concerning the ambiguities
in the definition 
of mass and width of relativistic quasi-stable particles, in
particular for the $Z$-boson \cite{PDG,Riemann, L3,Berends,
Berends.etal,Bardin.etal,Willenbrock.etal,Stuart,Sirlin},
we explore the
use of relativistic Gamow kets to provide unambiguous
definitions of mass and width.  The theoretical foundations of the
relativistic Gamow vectors are based on the time asymmetric Poincar\'{e}
transformations of relativistic spacetime and have recently been presented
elsewhere~\cite{preprint}.

\section{Lineshape Parameters and \\ Relativistic Breit-Wigners}
\label{LPandRBV}

The determination of the $Z$-boson mass and width from the lineshape
has been performed with two different definitions of mass and width
and two different relativistic Breit-Wigner amplitudes.  As a result, 
two different values for $m$ and $\Gamma$ have been obtained from the same
experimental data~\cite{PDG,Riemann,L3}.

The first approach, followed by practically all experimental analyses
of the LEP and SLC data~\cite{PDG}, is based on the on-shell
definition of mass and width, which we shall denote $M_Z$ and
$\Gamma_Z$.  The $Z$-boson mass and width are 
defined in perturbation theory by the (transverse) self-energy $\Pi(s)$ of the
$Z$-boson propagator.  The part of the $Z$-boson
propagator proportional to $\eta_{\mu\nu}$ is given by
\begin{equation}\label{propagator1}
D(s) = \frac{1}{s-M_0^2-\Pi(s)}.
\end{equation}
The on-shell scheme defines the (real) mass
$M_Z$ and the with $\Gamma_Z$ by the renormalization conditions:
\begin{subequations}\label{scheme1}
\begin{equation}
M_Z^2 = M_0^2 + \mathrm{Re} \Pi(M_Z^2),\ \ \ M_Z\Gamma_Z = 
\frac{-\mathrm{Im}\Pi (M_Z^2)}{1 - \mathrm{Re}\Pi'(M_Z^2)},
\end{equation}
where the prime indicates differentiation with respect to $s$.  Using
these conditions and expanding $\rm{Re}\Pi(s)$ about $s=M_Z^2$, $D(s)$
is written
\begin{eqnarray}
D(s) &=& \frac{1}{(s - M_Z^2) [ 1 - \mathrm{Re}\Pi'(M_Z^2)] - i
\rm{Im}\Pi(s)}\nonumber \\
&=& \frac{1}{1 - \mathrm{Re}\Pi'(M_Z^2)}\times\frac{1}{s - M_Z^2 +
i\sqrt{s}\Gamma_Z(s)}
\end{eqnarray}
\end{subequations}
where the ``$s$-dependent width'' $\Gamma_Z(s)$ has been defined by
\begin{equation}
\sqrt{s}\Gamma_Z(s) \equiv \frac{-\mathrm{Im}\Pi(s)}{1 -
\mathrm{Re}\Pi'(M_Z^2)} 
\end{equation}
This leads, to the ``relativistic Breit-Wigner with energy dependent
width'' [e.g.\ p.\ 189 of~\cite{PDG}] for the
lineshape of the $Z$-boson 
\begin{equation}
a_j^Z(s) = \frac{-\sqrt{s}\sqrt{\Gamma_e(s) \Gamma_f(s)}}{s - 
M_Z^2+i\sqrt{s}\Gamma_Z (s)}\label{s-dep.proto} 
\end{equation}
where the partial widths have the same $s$ dependence:
$\Gamma_{e,f}(s) = (\mbox{const.}) \times \Gamma_Z(s)$.  
Neglecting the fermion mass, one calculates for the $Z$-boson near the
resonance peak
\begin{equation}
\frac{\mathrm{Im}\Pi(s)}{1 -
\mathrm{Re}\Pi'(M_Z^2)} = -\frac{s}{M_Z} \Gamma_Z.
\end{equation}
Inserting this in (\ref{s-dep.proto}) (neglecting the irrelevant
$s$-dependence in the numerator), one obtains the standard expression
for the $j$th partial wave amplitude used in the line shape analysis
of all experiments
\begin{equation}
a_j^Z(s)\approx
\frac{-M_ZB_{ef}\Gamma_Z}{s-M_Z^2+i\frac{s}{M_Z}\Gamma_Z} = \frac{R_Z}{s -
M_Z^2+i\frac{s}{M_Z}\Gamma_Z}
\label{eq:s-dep.Breit-Wigner}
\end{equation}

The on-mass shell renormalization is
arbitrary~\cite{Berends.etal,Bardin.etal,
Sirlin,Stuart,Willenbrock.etal,Peskin.Schroeder} and may have some
problems with gauge invariance.  More serious may be the problem that
the conventional expressions of $\rm{Im}\Pi$ and therewith of $\Gamma$
treat the unstable state as an asymptotically free state, i.e.\ as
eigenvectors of the free Hamiltonian $H_0=H_f + H_{\bar{f}}$ and not
the full Hamiltonian $H = H_0 + V$, where $V$ is the decay
interaction, cf.\ (7.49) and (7.63) of~\cite{Peskin.Schroeder}.
Therefore another scheme based on the complex valued position of the
propagator pole $s_R = M_0^2 + \Pi(s_R)$ may be better because
scattering resonances are different from their asymptotically free in-
and out-states.

Writing the pole position as $s_R = M_\rho^2 - iM_\rho\Gamma_\rho$, one
obtains in place of (\ref{eq:s-dep.Breit-Wigner}) the expression
(\ref{eq:Breit-Wigner}) below which agrees phenomenologically with the
$S$-matrix definition.  The merit of the notation in terms of $M_R$
and $\Gamma_R$ will  become clear at the end of the paper.
 
In analytic $S$-matrix theory a resonance with spin $j$ is defined by a
(pair of) first order pole(s) at the position(s) $s_R =
(M_R-i\Gamma_R/2)^2$ (and $s_R^{*} =
(M_R+i\Gamma_R/2)^2$) on the second sheet of the analytically
continued $j$-th partial $S$-matrix element~\cite{Eden.etal}.
Analytic $S$-matrix theory 
defines the complex quantity $s_R$ in a model independent way.

Using the $S$-matrix definition of the $Z$-resonance (or the pole
definition of the propagator), the $Z$-boson pole
term of the $j$th partial amplitude is given by 
\begin{equation}
a_j^R(s) = \frac{R_Z}{s-M_\rho^2 + iM_\rho\Gamma_\rho} =
\frac{R_Z}{s-(M_R-i\frac{\Gamma_R}{2})^2} = 
\frac{R_Z}{s-s_R}  \label{eq:Breit-Wigner} 
\end{equation}
where $\Gamma_R$ and $M_R$ are basic $S$-matrix parameters and independent
of the energy $s$.  The same parameterization is obtained from the
complex pole definition of the $Z$-propagator.  We call
(\ref{eq:Breit-Wigner}) also a ``relativistic
Breit-Wigner'' or the $S$-matrix pole
Breit-Wigner\footnote{The ``relativistic
S-matrix pole Breit-Wigner''
(\ref{eq:Breit-Wigner}) in the non-relativistic limit becomes the
standard non-relativistic Breit Wigner
$(E-E_R+i\Gamma/2)^{-1}$.}. 

One can compare the two definitions of the mass $M_Z$ and $M_R$ by
adjusting the maximum $s_{max}^{(1)}$ of (\ref{eq:s-dep.Breit-Wigner})
and the maximum $s_{max}^{(2)}$ of (\ref{eq:Breit-Wigner}) to the peak position
of the experimental cross section data, $s_{peak}$.  The maximum of 
$|a_j^Z(s)|^2$ is $s_{max}^{(1)} =
M_Z^2(1+(\Gamma_Z/M_Z)^2)^{-1}$ and the maximum of
$|a_j^R(s)|^2$ is $s_{max}^{(2)} =
M_R^2(1-1/4(\Gamma_R/M_R)^2) \equiv M_{\rho}^2$ and from $s_{peak} =
s_{max}^{(1)} = s_{max}^{(2)}$ one obtains the
following differences in the values of $M_R$,
$M_{\rho}$ and
$M_Z$~\cite{Bardin.etal,Willenbrock.etal,Stuart,Sirlin}:
\begin{equation}
M_R^2 = M_Z^2 - \frac{3}{4}(\frac{\Gamma_Z}{M_Z})^2 M_Z^2 +
O((\frac{\Gamma_Z}{M_Z})^4) 
\end{equation}
or
\begin{equation}
M_R = M_Z - 26\mbox{MeV} \hspace{1.25in} M_{\rho} = M_Z - 34\mbox{MeV}.
\label{eq:mass.difference} 
\end{equation}

Both parameterizations (\ref{eq:s-dep.Breit-Wigner}) and
(\ref{eq:Breit-Wigner}) were fitted to the experimental cross sections
and asymmetries~\cite{Riemann, L3}.  These fits confirmed the
differences (\ref{eq:mass.difference}) and also
yielded $\Gamma_{\rho}-\Gamma_{Z} \approx (1-2)\mbox{MeV}$. 

The experimental fits incorporate more than just
(\ref{eq:Breit-Wigner}) or (\ref{eq:s-dep.Breit-Wigner}),
e.g. corrections, background and interference needed to be
incorporated into the analysis.  The pole term (\ref{eq:Breit-Wigner})
which enters into  the $j$-th partial amplitude $a_j(s)$ only
describes the part of the scattering which goes through the $Z$ resonance 
\begin{equation}\label{process}
e \overline{e} \rightarrow Z \rightarrow f \overline{f}.
\end{equation}
Even if there is only one intermediate particle with $j^P=1^-$ there is
\emph{always} a slowly varying background amplitude $B(s)$ so that
$a_j(s) = a^R_j(s) + B(s)$.  If there are
also other intermediate particles with $j^P=1^-$, for example only one
other with
(complex) mass squared $s=s_{R_2}$, then the partial amplitude contains a sum
of two BW's and a background term
\begin{equation}
a_j(s) = \frac{R_Z}{s-s_R} + \frac{R_2}{s-s_{R_2}} + B(s).
\label{eq:multiresonance.amplitude}
\end{equation}
This form (\ref{eq:multiresonance.amplitude}) does not follow from the
expansion of $a_j(s)$  into a Laurent series based on analyticity
assumptions about $a_j(s)$.  Instead it follows from a
stronger hypothesis of the
RHS formulation of quantum mechanics\footnote{\label{hypothesis} In
addition to the conventional analyticity
assumption (\ref{eq:multiresonance.amplitude}) requires also the
hypothesis that the prepared in-states $\phi^+$ and the observed
out-states (decay products) $\psi^-$ of the S-matrix element
$(\psi^-,\phi^+) = (\psi^{out},S\phi^{in})$ are
vectors of two different dense subspaces of the Hilbert Space ${\cal
H}$.  See Appendix.}.

The fit to the LEP data \cite{Riemann,L3} used for the amplitude an
expression which contained, in addition to the $Z$-boson Breit-Wigner
(\ref{eq:Breit-Wigner}) or (\ref{eq:s-dep.Breit-Wigner}), a  photon term
(``$\gamma$-Breit-Wigner'') and a slowly varying background amplitude
$B(s)$ which is assumed to be constant in the $Z$-boson energy
region~\cite{Leike.etal}:
\begin{equation}
a_j(s) = \frac{R_Z}{s-s_R} + \frac{R_{\gamma}}{s} + B(s),\
\mbox{with}\ s_R=(M_R - i\Gamma_R/2)^2.
\label{eq:total.amplitude}
\end{equation}
The external line QED corrections are included by a convolution
integral with the basic scattering cross sections given by
$|a_j(s)|^2$ of (\ref{eq:total.amplitude}).  The $\gamma$-Breit-Wigner
is the analogue of the one photon exchange 
graph of the internal line QED corrections.  Its inclusion in
(\ref{eq:total.amplitude}) treats the scattering as a double
multichannel resonance process~\cite{Bohm},
\begin{equation}
e\overline{e} \,\, \rightarrow \,\, \stackrel{Z}{_{\gamma}} \,\,
\rightarrow \,\, f\overline{f}. 
\end{equation}
In the $S$-matrix approach the amplitude (\ref{eq:total.amplitude})
may be considered as a limiting case of the double resonance amplitude
(\ref{eq:multiresonance.amplitude}) for $s_{R_2} 
\rightarrow 0$, though there is really no $S$-matrix theory
justification for (\ref{eq:total.amplitude}) except the analogy to
(\ref{eq:multiresonance.amplitude}).  However, the $Z$-$\gamma$
interference term from 
(\ref{eq:total.amplitude}) is important for the fits of various
asymmetries~\cite{Riemann}.

The earlier fits to the LEP data use the amplitude
with the $s$-dependent  denominator (\ref{eq:s-dep.Breit-Wigner}) from
the on-mass shell definition:
\begin{equation}
a_j(s) = \frac{R_Z}{s- M_Z^2 +is\frac{\Gamma_Z}{M_Z}} +
\frac{R_\gamma}{s} + B(s).\label{s.tot.amp}
\end{equation}
After the arbitrariness of this form became known, the same kinds of
fits were made using the $S$-matrix definition (\ref{eq:total.amplitude}).
The fits of the LEP data to (\ref{eq:total.amplitude}) and
(\ref{s.tot.amp}) were equally good and they reproduced the expected
difference (\ref{eq:mass.difference}) between the differently defined
masses and widths, $(M_Z, \Gamma_Z)$ from (\ref{s.tot.amp}) and $(M_R,
\Gamma_R)$ 
from (\ref{eq:total.amplitude}).  Thus the experimental lineshape data
of the $Z$-boson do not favor either of these two definitions of
$Z$-boson mass and width.

Recently the conventional
approach, using a Breit-Wigner with energy dependent width
(\ref{eq:s-dep.Breit-Wigner}), and the S-matrix approach, using the
pole definition (\ref{eq:Breit-Wigner}), were compared for hadron
resonances.  For baryon
resonances like the $\Delta_{33}$, recent editions of reference
\cite{PDG} list both the pole position
$s_h=(1210-i\frac{100}{2})^2\mbox{MeV}$ in addition to the
conventional parameters $M_h=1232\mbox{MeV}$ and $\Gamma(M_h^2)
=120\mbox{MeV}$.  The interpretation given to these
different values by \cite{Hohler.etc} is that the pole position $s_h$
belongs to the $\Delta$-resonance whereas the conventional parameters
($M_h$, $\Gamma(M_h^2)$) belong to the $\Delta$ and the background
together.  This interpretation is in agreement with our RHS theory of resonance
scattering as expressed by the results
(\ref{eq:multiresonance.amplitude}) and (\ref{eq:total.amplitude}).

When both the $S$-matrix definition (\ref{eq:Breit-Wigner}) and the
on-mass shell definition (\ref{eq:s-dep.Breit-Wigner}) were applied to the
$\rho$-meson data \cite{Bernicha.etal}, the conclusion was that the
S-matrix definition
of $M_{\rho}$ and $\Gamma_{\rho}$ is phenomenologically preferred.
The reason given was that these fitted parameters remained largely
independent of the parameterization of the background term $B(s)$ and
the $\rho-\omega$ interference.  A similar fit to the $S$-matrix Breit-Wigner
(\ref{eq:Breit-Wigner}) was performed for the experimental data on
$\pi p$ scattering in the $\Delta$ resonance
region~\cite{Bernicha.etal2}.  As with the $\rho$, the fits to
(\ref{eq:Breit-Wigner}) were better than the fits to
(\ref{eq:s-dep.Breit-Wigner}) and they were independent of the background
parameterization.  Also, the fitted values of $M_{\Delta}$ and
$\Gamma_{\Delta}$ from (\ref{eq:Breit-Wigner}) turn out to be significantly
smaller than the conventional values from
(\ref{eq:s-dep.Breit-Wigner}).  Therefore, for the hadron resonances
the pole definition (\ref{eq:Breit-Wigner}) may be phenomenologically
preferred.  Again, as stated above, the same  conclusion does not follow from
the $Z$-boson data 
\cite{Riemann,L3} where these two approaches lead just to different
values of the $Z$-boson parameters.

\section{Lifetime Parameter and \\ Exponential Decay Rate}

Neither the treatment of the resonant propagator based on the complex
pole position $s_R = M_0^2 + \Pi(s_R)$ nor the phenomenologically
equivalent $S$-matrix theory definition of a quasi-stable
particle by the pole at $s_R$ define the mass and width
separately.  A wide assortment of two real numbers can be extracted
from $s_R$ and 
identified with the mass and width~\cite{Stuart2}, e.g.~one could
write
\begin{equation}
s_R = (M_R-i\frac{\Gamma_R}{2})^2 = M_{\rho}^2 - iM_{\rho}\Gamma_{\rho}
\end{equation}
and call either $M_R$, $\Gamma_R$ or $M_{\rho}$, $\Gamma_{\rho}$ the
resonance mass and width.  There is no principle in analytic
$S$-matrix theory that tells us how to separate the complex number
$s_R$ into the real mass and width.  Lineshape by itself, however
accurately determined and however precisely the background and
corrections are known, does not determine mass and width separately.
However, there is a separate physical meaning for the width (and the
resonance mass).  For this we turn to the second
definition of $\Gamma$ given by the inverse lifetime $\Gamma =
\hbar/\tau$, as mentioned in section 1. 

The measurement of the lifetime uses the exponential law for the
partial decay rates $\dot {\cal P}_{\eta}(t)$ of the quasi-stable
particle $R$ into any decay products with the channel quantum numbers
$\eta$.  It is done by a fit of the experimental
counting rate $\dot N_\eta(t)/N = \dot {\cal P}_{\eta}(t)/\Gamma$
to the exponential $\exp(-t/\tau)$, where $t$ is the time in the
center of mass frame of the decay products $\eta$.  This cannot be
done for the $Z$-boson because $\hbar/\Gamma_Z$ would be too
short to measure, although it has been done for many other decaying
elementary particles like $\mu^{\pm}$, $\pi^{\pm}$, $K^{\pm,0}$.

All relativistic quasistable particles should have a width $\Gamma$ defined by
the Breit-Wigner and a lifetime $\tau$ defined by the exponential decay law.
Although both $\Gamma$ and $\tau$ may not be measured for the same
particle due to experimental limitations, we want to require that a
complete theoretical 
description of a resonance should provide an unambiguous, \emph{exact}
connection between them: the relation $\Gamma=\hbar/\tau$ should be fulfilled
exactly and universally.

To establish the width $\Gamma$ of (\ref{eq:Breit-Wigner}) as the
inverse lifetime we have to first establish
\begin{equation}
\dot {\cal P}_{\eta} (t) = \Gamma^{(\eta)} e^{-\Gamma_R t}
\label{eq:decay.rate}
\end{equation}
as a precise result obtained from the second sheet $S$-matrix pole
definition of a
resonance.  Then, the relationship $\Gamma = \hbar/\tau$ will hold
precisely and not just as a result of the WW approximate
methods~\cite{Goldberger.Watson,Weisskopf.Wigner}.  So we need an
intermediary that connects the Breit-Wigner (\ref{eq:Breit-Wigner}) to
the exponential law (\ref{eq:decay.rate}).  This will be the ``state
vector'' of the resonance state, the relativistic Gamow vector. 

Conventionally state vectors representing decaying states and
resonances are obtained using
heuristic finite dimensional models with an $n$-dimensional complex ``effective
Hamiltonian matrix'' $H^{eff}$.  For example, in the two dimensional
neutral Kaon system \cite{Lee.etal}, the elementary particle state is
defined as the eigenvector of the complex
Hamiltonian matrix with a complex eigenvalue 
\begin{equation}
H^{eff} f^{K_{S,L}} = (m_{S,L} - i\frac{\Gamma_{S,L}}{2})
f^{K_{S,L}}. \label{eq:KAONS} 
\end{equation}
From this one concludes
\begin{equation}
f^{K_{S,L}}(t) \equiv e^{-iH^{eff}t} f^{K_{S,L}} = e^{-im_{L,S}t}
e^{-\frac{\Gamma_{S,L}t}{2}} f^{K_{S,L}}   \label{eq:kaons} 
\end{equation}
and the general state vector is taken as the superposition
\begin{subequations}\label{kstate}
\begin{equation}
\phi = c_sf^{K_S} + c_Lf^{K_L}\label{Katt0}
\end{equation}
with the time evolution
\begin{eqnarray}
\phi(t) &=& e^{-H^{eff}t}\phi = c_S f^{K_S}(t) + c_L f^{K_L}(t)\nonumber\\
&=& c_Se^{-im_St}e^{-\Gamma_St/2}f^{K_S} +
c_Le^{-im_Lt}e^{-\Gamma_Lt/2}f^{L_S}.  \label{eq:general.kaon.state}
\end{eqnarray}
\end{subequations}
The exponential law (\ref{eq:decay.rate}) is then justified in the
following way: one inserts (\ref{eq:kaons}) for all values of $t$ into
Dirac's Golden Rule and obtains a rate $R(t)$ for all values of
$t$\footnote{Additionally, one inserts
(\ref{eq:general.kaon.state}) into Dirac's
Golden Rule and obtains the decay rate for the neutral Kaon
state considered as a superposition of two exponentially decaying
states, but this matter is not the subject of the present paper.}.
This $R(t)$ one then equates with the time derivative of the decay
probability $\dot{\mathcal{P}}_\eta(t)$.
This deduction of the exponential law has the following deficiencies:
\begin{enumerate}
\item Dirac's Golden Rule
is only an approximation for the initial decay rate (i.e.\ for $t
\rightarrow 0$) and has only been obtained for the
Dirac Lippmann-Schwinger 
eigenkets of real energy $|E^- \rangle$ and not for vectors like
$f^{K_S}$ with complex energies
($E-i\Gamma/2$).  To justify (\ref{eq:decay.rate}) one should start 
from the fundamental
quantum mechanical formula for the probability of decay from $R$ to
$\eta$
\begin{equation}\label{eq:decay.prob}
{\cal P}_{\eta}(t)={\rm Tr}(\Lambda_\eta|\psi^{\rm
G}(t)\rangle\langle\psi^{\rm G}(t)|),
\end{equation}
where $\psi^G$ describes the
decaying state $R$ and $\Lambda_{\eta}$ is the projection operator on
the space of decay products $\eta$.  Then the probability rate
(\ref{eq:decay.rate}) should result as the time derivative of ${\cal
P}_{\eta}(t)$~\cite{Harshman}.

\item $\Gamma_S$ is defined by (\ref{eq:KAONS}) as the imaginary
part of a complex energy and from (\ref{eq:kaons}) one sees this is thus
just another name for $\hbar/\tau$.  It is unrelated to the width of a
lineshape (\ref{eq:Breit-Wigner}) because (\ref{eq:KAONS}) and its
underlying assumptions do
not imply a precise relation between $\Gamma_S$ and any width $\Gamma$
(either $\Gamma_R$, $\Gamma_\rho$, $\Gamma_Z$ or others) of any
relativistic Breit-Wigner.  To justify $\tau=\hbar/\Gamma$ one must
relate $\Gamma_S$ of (\ref{eq:KAONS}) to the $\Gamma_R$ (or
$\Gamma_{\rho}$, etc.~$\!\!$) of the pole definition of the lineshape
(\ref{eq:Breit-Wigner}) or any
other definition for the lineshape, e.g.\ (\ref{eq:s-dep.Breit-Wigner}).

\item The energy operator $H$ (or in the rest system, $H \equiv P_0 = M =
(P_\mu P^\mu)^{1/2}$ for the relativistic case like the $K$-meson) is a
self-adjoint operator in an infinite dimensional space, usually the
representation 
space of the Poincare group of transformations of spacetime.  To
justify the two dimensional ``complex'' eigenvector expansion
of a general state vector $\phi$ (\ref{Katt0}), one
first has to show that the eigenvectors with complex eigenvalue
(\ref{eq:KAONS}) have a meaning for self-adjoint $H$, that they can be
elements of a basis system for the infinite dimensional space and that
the truncation to the two-dimensional subspace gives 
(\ref{Katt0}) as an approximation.
\end{enumerate}

All these things can be justified, with some qualifications, 
within the RHS formulation of time
asymmetric quantum 
theory by augmenting the conventional assumptions of scattering theory
with the new hypothesis described in the Appendix.

From this new hypothesis one derives both the form of the $j$th
partial amplitude 
(\ref{eq:multiresonance.amplitude}) for two resonance poles of the
partial $S$-matrix and a complex
basis vector decomposition for the prepared in-state
$\phi^+$, i.e.\ $e\bar{e}$ in case of (\ref{process}), given by (cf.\
eq.\ (6.24) of \cite{Harshman} for the 
non-relativistic case):
\begin{eqnarray}
\phi^+ & = & \sum_{R_i} |j,s_{R_i};b_i {}^- \rangle c_{R_i} + \phi^{bg}
       \nonumber \\  
       & = & |j=1^-,s_R;b_R {}^- \rangle \label{eq:basis.decomposition}
       c_R + |j=1^-, s_{R_2};b_{R_2} {}^- \rangle c_{R_2} + \phi^{bg},
\end{eqnarray}
which is reminiscent of (\ref{Katt0}).
Here $b$ denotes a set of degeneracy labels\footnote{For $b$ one could
choose the momentum $\mathbf{p}_m$ and $j_3$ but we suggest the
spatial components of the 4-velocity $\hat{\mathbf{p}}_m =
\mathbf{p}_m/s$.  However this is not important for our discussions here;
for a detailed discussion see~\cite{preprint}.}  

The terms in the basis vector decomposition
(\ref{eq:basis.decomposition}) correspond to the terms in
the multi-resonance partial amplitude
(\ref{eq:multiresonance.amplitude}).  To each
vector in the sum over the resonances $R_i$ in
(\ref{eq:basis.decomposition}) corresponds a BW in the $j$th partial
amplitude (\ref{eq:multiresonance.amplitude}).  The vector
$\phi^{bg}$ is the 
component of the (arbitrary) in-state vector $\phi^+$ in
the infinite dimensional subspace representing the non-resonant
continuum and it corresponds
to the $B(s)$ term in (\ref{eq:multiresonance.amplitude}).  In the
next section we will show how to obtain the decomposition
of the in-state into a sum over resonances and a non-resonant
part by the
analytic continuation of the partial $S$-matrix.  The
vector $|j,s_{R_i};b^-\rangle$ in the sum over $R_i$ in
(\ref{eq:basis.decomposition}) 
comes from the pole at $s_{R_i}$ in the second sheet of the $S$-matrix and the
vector $\phi^{bg}$ is represented by a background
integral (see (\ref{eq:S-matrix.background}) below).    It is
because we can obtain both the BW amplitude (\ref{eq:Breit-Wigner})
and the resonance vector $\mkt{j,s_R;b}$ from the $S$-matrix pole that
we can exactly and precisely establish the
relation $\tau=\hbar/\Gamma_R$.

Before continuing with the details of the derivation in the following
section, a few comments are in order about the vectors \mkt{j,s_R;b},
which (except for a normalization factor) are the relativistic Gamow
vectors $\psi^G$.

For stable relativistic elementary particles the vector space description
is given by the irreducible unitary representation spaces of the
Poincare group ${\cal P}$~\cite{Wigner}, from which we can then define
the fields.  This is not restricted to the asymptotic,
interaction free states~\cite{Weinberg}.
The `out' and `in' ``states'' of \cite{Weinberg}, which we denote by
$\mkt{j,s;b}$ and $\pkt{j,s;b}$ and which fulfill the
Lippmann-Schwinger equation, are basis vectors of an
irreducible unitary representation $(j,s)$ of ${\cal
P}$.  They are eigenkets of the
(self-adjoint) invariant mass squared operator $P_{\mu} P^{\mu} \equiv
M^2$ with real eigenvalue $s$:
\begin{eqnarray}
P_{\mu} P^{\mu} |j,s;b^-\rangle & = &
s|j,s;b^-\rangle  \label{LSm2}  \\ 
P_0|j,s;b=b_{rest}^-\rangle & = & \sqrt{s}|j,s;b=b_{rest}^-\rangle.
\label{LSp0} 
\end{eqnarray}
We want to consider the $Z$-boson (or any
quasistable relativistic particle) as a fundamental particle in the
Wigner sense and therefore give its description in terms of a
representation space of Poincar\'{e} group transformations.  The Gamow
vectors $\mkt{j,s_R;b}$ of (\ref{eq:basis.decomposition}) should
therefore also be basis vectors of an irreducible representation
$(j,s_R)$ of Poincar\'{e} transformations.  The Gamow kets
$\mkt{j,s_R;b}$ are thus just generalizations of the
well-established `out-states' $\mkt{j,s;b}$ and are obtained from them
by analytic 
continuation in $s$ from its ``physical value'' $\{s|(m_e +
m_{\bar{e}})^2 \leq s < \infty\}$ the the $S$-matrix pole position
$s_R$ in the second sheet (or if there are several poles, a Gamow ket 
$\mkt{j,s_{R_i};b}$ is obtained for each pole $s_{R_i}$).  In place of
(\ref{LSm2}) and (\ref{LSp0}) for the Dirac Lippmann-Schwinger kets
$\pmkt{j,s;b}$, the Gamow kets should be generalized eigenvectors of
the self-adjoint mass-squared operator $P_\mu P^\mu\equiv M^2$ with
complex eigenvalue $s_R = (M_R + i\Gamma_R/2)^2$:
\begin{eqnarray}
P_{\mu} P^{\mu} |j,s_R;b^-\rangle & = &
s_R|j,s_R;b^-\rangle  \label{eq:msquared}  \\ 
P_0|j,s_R;b=b_{rest}^-\rangle & = & (M_R -
i\frac{\Gamma_R}{2})|j,s_R;b=b_{rest}^-\rangle.
\label{eq:action.of.P0} 
\end{eqnarray}
These eigenvalue equations are exact and mathematically rigorous if the
$\mkt{j, s_R; b}$ are defined as continuous functionals over the space of
observables $\Phi_+$ (cf. Appendix (\ref{eq:bc})).  More details will
be given in the following section, but for further details
see~\cite{preprint}.

The eigenvalue equation
(\ref{eq:action.of.P0}) agrees with the complex Hamiltonian
eigenvalue equation 
(\ref{eq:KAONS}) and the decomposition
(\ref{Katt0}) is the truncation of
(\ref{eq:basis.decomposition}) to 
the resonance subspace if the background $\phi^{bg}$ (or equivalently
$B(s)$ in (\ref{eq:multiresonance.amplitude})) is neglected.  Thus the
Lee-Oehme-Yang theory~\cite{Lee.etal} of the neutral kaon system
summarized by (\ref{eq:KAONS}-\ref{kstate}) and other
finite dimensional effective theories with complex Hamiltonians
\cite{Ferreira} will be established as approximations of the RHS
theory after (\ref{eq:basis.decomposition}, \ref{eq:msquared},
\ref{eq:action.of.P0}) have 
been justified in the next section.

\section{Relating the BW Lineshape \\ to the Exponential Decay Law: \\
Relativistic Gamow Vectors}

The mathematical
object that makes the connection between the $\Gamma$ that appears in
the lineshape  
(\ref{eq:Breit-Wigner}) and the $\Gamma$ that appears in
the exponential decay law (\ref{eq:decay.rate}) is the relativistic
Gamow vector 
$\psi^G$.  The vector $\psi^G$ fulfills (\ref{eq:msquared}) and
(\ref{eq:action.of.P0}) 
precisely and is obtained from the pole term of the $S$-matrix.  The
Gamow vector $\psi^G$ will on the one hand lead to the lineshape
(\ref{eq:Breit-Wigner}) and on the other hand to a precise form of the
exponential law (\ref{eq:kaons}).  Since both come from the same vector
characterized by $\Gamma_R$, the lineshape-$\Gamma$ and the inverse
lifetime-$\Gamma$ are the same and given by $\Gamma_R$.  Because the
Gamow vector distinguishes $\Gamma_R$ as the characterizing parameter,
it implies a preferred separation of the complex value $s_R$ into two
real parameters $(M_R,\Gamma_R)$: 
$s_R=(M_R - i\Gamma_R/2)^2$.

We start with the $S$-matrix element
\begin{eqnarray}
(\psi^{out},S \phi^{in}) &=& (\psi^-,\phi^+)\nonumber \\
& =& \sum_{j}
\int_{(m_e+m_{\overline{e}})^2}^{\infty}\!\!\!\!\!\!\! ds \sum_{b} \langle
\psi^-|j,s;b^- \rangle S_j(s) \langle ^+ j,s;b|\phi^+ \rangle,
\label{eq:S-matrix} 
\end{eqnarray}
where $\phi^+ \in \Phi_-$ represents the prepared in-state (e.g.\ $e^+e^-$) and
$\psi^- \in \Phi_+$ represents the detected out-state (e.g.\ decay
products $f\bar{f}$).
Then to obtain the Gamow vectors one proceeds in exactly the same way as in
the non-relativistic case~\cite{Bohm.etal}.  One deforms the contour
of integration from the upper rim of the cut along the positive real
axis $(m_e+m_{\overline{e}})^2 \leq s < \infty$ through the cut into
the second (or higher) sheet past the pole at $s_R$ (or past the poles
at $s_R$,$s_{R_2},\ldots$ if there is more than one resonance pole).
For the contour around each resonance pole $s_R$, the integral in
(\ref{eq:S-matrix}) splits off a pole term which gives the Gamow
vector $|j,s=s_R;b^- \rangle$ as an analytic continuation of the
Dirac-Lippmann-Schwinger kets $|j,s;b^- \rangle$.  The Gamow kets are defined
(using the Cauchy formula) by the contour integral around the
resonance pole $s_R$
\begin{subequations}\label{eq:S-matrix.pole}
\begin{eqnarray}
\langle \psi^-|j,s_R;b^- \rangle &\equiv& -\frac{i}{2\pi} \oint ds
\langle \psi^-|j,s;b^-\rangle \frac{1}{s-s_R} \label{poleA}\\
& = &\frac{i}{2\pi}
\int_{-\infty_{II}}^{+\infty} ds \langle \psi^-|j,s;b^-\rangle
\frac{1}{s-s_R}.  \label{poleB} 
\end{eqnarray}
\end{subequations}
The integral in (\ref{eq:S-matrix}) thus becomes a sum of pole terms
(one for each $s_{R_i}$) plus a background integral 
\begin{equation}
\int_{\cal C} ds \langle \psi^-|j,s;b^-\rangle S_j(s) \langle ^+
j,s;b|\phi^+ \rangle \equiv \langle \psi^-|\phi^{bg} \rangle
\label{eq:S-matrix.background} 
\end{equation}
over a contour ${\cal C}$, which we are largely free to choose far
away from the resonance poles (e.g. along the negative real
axis on the second sheet).  In this way one arrives at the
representation (\ref{eq:multiresonance.amplitude}) for the $j$-th
partial amplitude (or similarly for the $j$-th partial S-matrix
$\langle b'|S_j(s)|b \rangle=S_j(s)$).  By omitting the arbitrary
$\psi^-\in\Phi_+$ in (\ref{eq:S-matrix}) one arrives at the complex
basis vector decomposition (\ref{eq:basis.decomposition}) for
the in-state vector $\phi^+$ where $\phi^{bg}$ is given by
(\ref{eq:S-matrix.background}) with the arbitrary $\psi^- \in \Phi_+$
omitted.  Summarizing, we have obtained by contour deformation the BW
amplitude (\ref{eq:multiresonance.amplitude}) and the basis vector
decomposition 
(\ref{eq:basis.decomposition}) from the $S$-matrix element and
established the correspondence between the terms of these equations.
From (\ref{poleB}) we see that the variable $s$ in
(\ref{eq:Breit-Wigner}), (\ref{eq:multiresonance.amplitude}), etc.\,
extends from $-\infty_{I\!I}$ to $+\infty$; however for ``unphysical'' values
of $s\leq (m_e + m_{\bar{e}})^2= 4m_e^2$ these values are in
the second sheet.

In order to perform the contour deformation that separates
(\ref{eq:S-matrix}) into a sum over resonant terms like
(\ref{eq:S-matrix.pole}) plus the background term
(\ref{eq:S-matrix.background}) and in order to derive the Breit-Wigner
energy distribution of (\ref{poleB}) from the pole in (\ref{poleA})
some mathematical properties must be fulfilled in addition to the
conventional assumptions.  This is the new hypothesis of the Appendix
and here means specifically that the energy wave functions $\langle ^-
j,s;b|\psi^- \rangle$ and $\langle ^+ j,s;b|\phi^+ \rangle$ must be
well-behaved Hardy class functions\footnote{The proof of the second
equality in (\ref{eq:S-matrix.pole}) and of relations
(\ref{eq:action.of.M2}) and (\ref{eq:action.of.H}) is a consequence of
the following Titchmarsh theorem for Hardy class functions $G_-(s) \in
{\cal H}_-^2$:

For $G_-(s) \in {\cal H}_-$ (Hardy class in the lower plane) and
$\rm{Im}\, z > 0$ one has
\begin{equation}
G_-(z) = -\frac{1}{2\pi i} \int_{-\infty}^{+\infty} \frac{G_-(s)}{s-z}
ds \nonumber
\end{equation}
The functions $\bk{^+s}{\phi^+}$,$\langle \psi^-|s^- \rangle \in {\cal
H}_-$ are \emph{well-behaved} Hardy-class functions in the lower half
plane and $\bk{\phi^+}{s^+}$,$\langle ^-s|\psi^- \rangle \in {\cal
H}_+$ are well-behaved Hardy-class functions in the upper half plane.
This means that $\sqrt{s} \langle \psi^-|s^- \rangle, s \langle
\psi^-|s^- \rangle, \ldots, s^{\frac{n}{2}} \langle \psi^-|s^-
\rangle, \ldots$ (for $n=0,1,2,3,\ldots$) are also well-behaved
elements of ${\cal H}_-$.  Choosing $G_-(s) = s^{\frac{n}{2}} \langle
\psi^-|s^- \rangle$ one obtains (\ref{eq:S-matrix.pole}),
(\ref{eq:action.of.M2}), (\ref{eq:action.of.H}) and
more. \label{Hardy.footnote}} of the upper and lower half-plane,
respectively, in the second sheet of the energy surface of the
S-matrix, e.g. $\langle ^-s|\psi^- \rangle \in \mathcal{S}\cap{\cal
H}^2_+$.

The relativistic Gamow vectors $\mkt{j, s_R, b}$ defined by
(\ref{eq:S-matrix.pole}) satisfy the eigenvalue equations
\begin{eqnarray}
\langle \psi^-|(M^2)^{\times}|j,s_R;b^- \rangle &=& \frac{i}{2\pi}
\int_{-\infty}^{+\infty} ds\, s \langle \psi^-|j,s;b^- \rangle
\frac{1}{s-s_R}\nonumber\\ & =& s_R \langle \psi^-|j,s;b^- \rangle\,
\label{eq:action.of.M2} 
\end{eqnarray}
for every $\psi^- \in \Phi_+ \subset \HS \subset \Phi_+^\times$.  To
prove (\ref{eq:action.of.M2}) one  needs to use the properties of
Hardy class spaces discussed in footnote \ref{Hardy.footnote}.
Similarly, one can show that the Gamow vectors in the rest frame
$|j,s_R;b^- \rangle \rightarrow  |j,s_R;b_{rest}{}^- \rangle$ are
generalized eigenvectors of the energy operator $H=P_0$
\begin{eqnarray}
\langle \psi^-|H^{\times}|j,s_R;b_{rest}{}^- \rangle & = &
 \frac{i}{2\pi} \int_{-\infty}^{+\infty} ds \sqrt{s} \langle
 \psi^-|j,s;b_{rest}{}^- \rangle \frac{1}{s-s_R} \nonumber \\  & = &
 (M_R-i\frac{\Gamma_R}{2}) \langle \psi^-|j,s_R;b_{rest}{}^- \rangle
 \label{eq:action.of.H},
\end{eqnarray}
where $\sqrt{s_R} = (M_R-i\frac{\Gamma_R}{2})$.  The eigenvalue
equations (\ref{eq:action.of.M2}) and (\ref{eq:action.of.H}) are
precise formulations of (\ref{eq:msquared}) and
(\ref{eq:action.of.P0}).  The operators $H^{\times}$   and
$(M^2)^{\times}$ denote the conjugate operators in $\Phi_+^{\times}$
of the operators $H$ and $M^2 \equiv P_{\mu}P^{\mu}$ in the space
$\Phi_+$ and are defined by $\langle H \psi^-|F^- \rangle \equiv
\langle \psi^-|H^{\times}|F^- \rangle$ and $\langle M^2 \psi^-|F^-
\rangle \equiv \langle \psi^-|M^{2^{\times}}|F^- \rangle$ for all
$\psi^- \in \Phi_+$ and $F^- \in \Phi_+^{\times}$ (see
\cite{Bohm.etal} for more on their definition and use).

The Gamow kets $\mkt{j,s_R;b}$  are also generalized eigenvectors with
generalized eigenvalues $b=b_1,b_2,\cdots$ in the sense of Dirac kets.
Continuous linear combinations of the Gamow kets with an arbitrary
weight function $\phi_{j_3}(b) \in {\cal S}$ (Schwartz space)
\begin{equation}\label{super}
\psi^G_{j s_R} = \sum_{j_3}\int d\mu(b) \mkt{j,s_R;b} \phi_{j_3}(b),
\end{equation}
also represent relativistic Gamow states with complex mass $\sqrt{s_R}
= (M_R-i\Gamma_R/2)$ and a Breit-Wigner energy distribution $\langle
^- j,s;b|\psi_{j s_R} \rangle \propto (s-s_R)^{-1}$.

For the time evolution we consider here only the Gamow states at rest,
where $t$ is the proper time.  We calculate for all $\psi^- \in
\Phi_+$
\begin{eqnarray}\label{timeevol}
\langle e^{iHt}\psi^-|s_R;b_{rest}{}^-\rangle & = & \langle
	\psi^-|e^{-iH^{\times}t}|s_R;b_{rest}{}^- \rangle \nonumber \\
	& =& \frac{i}{2\pi} \int ds \frac{\langle
	\psi^-|e^{-iH^{\times}t}|s;b_{rest}{}^- \rangle}{s-s_R}
	\nonumber \\  & = & \frac{i}{2\pi} \int ds
	\frac{e^{-i\sqrt{s}\,t}\bk{\psi^-}{s;b_{rest}{}^-}}{s-s_R}\nonumber
	\\ &=& e^{-i\sqrt{s_R}t} \bk{\psi^-}{s_R;b_{rest{}^-}}
\end{eqnarray}
The last equality holds iff $\bk{\psi^-}{s^-} \in {\HS_-}$
\emph{and} $e^{-i\sqrt{s}t}\bk{\psi^-}{s^-} \in\HS_-$ because only then can
one also use the Titchmarsh formula of footnote \ref{Hardy.footnote} for
$G_-(s) = e^{-i\sqrt{s}t}\bk{\psi^-}{s^-}$ ($\pi < \mbox{arg}\sqrt{s}
\leq 2\pi$) and obtain $G_-(s_R)$. This is fulfilled only for $t \geq
0$ and not for $t<0$.  Thus, due to the new hypothesis of the Appendix,
we have exponential time evolution, but only for $t \geq 0$.  The dual
operator $(e^{iHt})^{\times} \equiv 
e^{-iH^{\times}t}$ of $e^{iHt}$ is not defined for $t<0$ because $e^{iHt}$ is
not a continuous operator in $\Phi_+$.  Omitting the arbitrary
$\psi^-\in\Phi_+$ in (\ref{timeevol}), we obtain the time evolution of
the relativistic
Gamow states at rest:
\begin{equation}
e^{-iH^{\times}t} \mkt{j,s_R;b_{rest}} = e^{-iM_R t}e^{-\Gamma_R t/2}
\mkt{j,s_R;b_{rest}}, \,\, t\geq 0.
\label{eq:time.evolution.Gamow.vector} 
\end{equation}
This is a functional equation over the space $\Phi_+ = \{\psi^-\}$,
describing the detected decay products.  This means it only makes
mathematical sense when used to calculate the probabilities $|\langle
\psi^-|e^{-iH^{\times}t}|j,s_R \rangle|^2$ of the decay products
$\psi^-$; the ``scalar products'' of
(\ref{eq:time.evolution.Gamow.vector}) with $\phi^+$ or $\psi^G_{j
s_R}$ are not defined.  This is as far as we can establish the
heuristic equation (\ref{eq:kaons}) by the mathematically rigorous
result (\ref{eq:time.evolution.Gamow.vector}).  That it holds for
$t\geq 0$
only does not constitute a limitation for the
physics.  Just to the contrary, $t\geq0$ describes the physical
situation correctly because the decay products $\eta$, described by
$\psi^-$, can only be detected \emph{after} the decaying state $R$, described
by the Gamow vector $|j,s_R^- \rangle$, has been created at
$t=0$~\cite{Bohm2}.

The time evolution (\ref{eq:time.evolution.Gamow.vector}) holds for
every Gamow vector in the basis vector expansion
(\ref{eq:basis.decomposition}).  Therewith the time evolution of the
heuristic state vector (\ref{eq:general.kaon.state}) as a
superposition of two exponentials can also be justified for $t\geq0$
if $\phi^{bg}$ in (\ref{eq:basis.decomposition}) can be neglected.
While the time evolution of the resonance terms in
(\ref{eq:basis.decomposition}) depends only upon the parameters
($M_{R_i}$, $\Gamma_{R_i}$) and is exponential, the time evolution of
the non-resonant 
background $\phi^{bg}$ is non-exponential and depends upon the
particular choice of the prepared in-state $\phi^+$ as seen from
(\ref{eq:S-matrix.background}).  The time evolution of
$|\bk{\psi^-}{\phi^+(t)}|^2$ can be very close to exponential if
$\phi^+$ is prepared such that $\phi^{bg} \approx 0$ \cite{Martin} (and
there is only one resonance in (\ref{eq:basis.decomposition})).

\section{Summary and Conclusion}

The relativistic Gamow vector
has the exact exponential time evolution
(\ref{eq:time.evolution.Gamow.vector}) and the exact S-matrix pole
Breit-Wigner energy distribution (\ref{eq:Breit-Wigner}).
Since these are the signatures of a quasistable particle and a
relativistic resonance, we want to assign the relativistic Gamow vectors
as the ``state vectors'' of quasistable relativistic particles.
On the basis of
(\ref{eq:time.evolution.Gamow.vector}) we want $M_R$ and $\Gamma_R$ to
define ``mass'' and ``width'' of a quasistable
relativistic particle, and from (\ref{eq:time.evolution.Gamow.vector})
it follows that the lifetime is exactly $\tau=\hbar/\Gamma_R$ and not
$\hbar/\Gamma_Z$ or $\hbar/\Gamma_\rho$.

Since the resonance always occurs with at least some background,
described by $B(s)$ of the amplitude
(\ref{eq:multiresonance.amplitude}) and by the much lesser known
(because it is standardly ignored, e.g.\ in
(\ref{kstate})) $\phi^{bg}$ of the prepared state
$\phi^+$, one may doubt the utility of the definition of a Gamow
state vector.  However, stable
elementary particles also never occur in  total isolation and the
accuracy with which the exponential law has been observed in some
cases \cite{exponential.fit} shows that the isolation of a
microphysical Gamow state $\psi^G$ from the background $\phi^{bg}$ can
be very good.  The popularity of the effective theories with finite
dimensional complex Hamiltonian matrices not only in particle physics
\cite{Lee.etal} but also in other areas~\cite{Ferreira} testifies to
the usefulness of separating exponentially evolving state vectors.
The relativistic Gamow vector is not more nor less of an idealization
of reality than Wigner's unitary representations for stable particles.
A vector description (or in general a density operator
description) is needed because the fundamental probabilities
(\ref{eq:decay.rate}) and (\ref{eq:decay.prob}) of quantum mechanics
are calculated in terms of operators or vectors.  To define a quantum
physical entity entirely by the $S$-matrix pole alone would be incomplete.

If the description of resonances by relativistic Gamow vectors is
valid, then the lineshape (\ref{eq:Breit-Wigner}) with mass (and by
(\ref{eq:time.evolution.Gamow.vector})) resonance energy in the rest
frame) given by $M_R=91.1626\pm .0031\mbox{GeV}$ and the width (by
(\ref{eq:time.evolution.Gamow.vector}) the inverse lifetime) given by
$\Gamma_R=2.4934\pm .0024\mbox{GeV}$ is its first prediction.  This
differs from the 
conventional mass definition by the lineshape (\ref{eq:s-dep.Breit-Wigner})
of the on-mass shell renormalization scheme $M_Z=91.1871 \pm
.0021\mbox{GeV}$ and it also differs from the definition by
 the peak position of the relativistic BW (\ref{eq:Breit-Wigner}),
$M_\rho=91.1541\pm .0031\mbox{GeV}$~\cite{CERN}.

This prediction was obtained, like the relativistic Gamow vector, from
the definition of the resonance as a pole of the $S$-matrix at $s_R$,
which by itself is insufficient to fix $M_R=\mathrm{Re}\sqrt{s_R}$ and
$\Gamma_R = -2\mathrm{Im}\sqrt{s_R}$.  Fixing the mass and width could
only be done using 
the new hypothesis of time asymmetric quantum physics, (\ref{eq:bc})
and (\ref{innout}) of the Appendix.

The other results of this theory like
the exponential law with the precise $\tau = \frac{\hbar}{\Gamma_R}$,
the basis vector expansion (\ref{eq:basis.decomposition}) or its
truncation to the complex ``effective'' theories like
(\ref{eq:KAONS}-\ref{eq:general.kaon.state}), the
representation (\ref{eq:multiresonance.amplitude}) of interfering
Breit-Wigners associated to (\ref{eq:basis.decomposition}) all have been
introduced before as separate assumptions and it may be welcome that
here they all follow from the same new hypothesis about the boundary
conditions.

The only result which may be difficult to accept is the semigroup
property of the time evolution (\ref{eq:time.evolution.Gamow.vector}).
In the relativistic theory this means that Gamow vectors can only
undergo Poincar\'{e} transformations into the forward light
cone~\cite{preprint}.  These  Poincar\'{e} transformations form only a
semigroup ${\cal P}_+=\{ (a, \Lambda)\}$, where $\Lambda$ is a
proper orthochronous Lorentz transformation and $a = (a_0,
\mathbf{a})$ is a four vector which fulfills $\hat{p}\cdot a = \sqrt{1
+ \mathbf{\hat{p}}^2}a_0 - \mathbf{\hat{p}}\cdot \mathbf{a}\geq 0$ for
any $\mathbf{\hat{p}}\in\mathbb{R}^3$\footnote{One checks by direct
calculations that $(a_1,\Lambda_1)\circ(a_2, \Lambda_2) = (a_1 +
\Lambda_1 a_2, \Lambda_1\Lambda_2)\in {\cal P}_+$ for $(a_i,
\Lambda_i)\in{\cal P}_+$ but that not every $(a, \Lambda)\in {\cal
P}_+$ has an inverse in ${\cal P}_+$, i.e.~${\cal P}_+$ is a
semigroup.}.

This semigroup property was a surprising and unintended result when it
was first derived for the non-relativistic theory.  
It expresses a time asymmetry on the microphysical level which is
connected with neither the violation of time reversal invariance of the
Hamiltonian nor entropy increase (at least not in an obvious way,
although for a contrary opinion see~\cite{G-M}).
In the meanwhile,
the irreversible character of quantum mechanical decay has been mentioned in
a few textbooks~\cite{Cohen} and more general considerations support the
existence of a fundamental time asymmetry in the quantum theory of
cosmology~\cite{G-M} and of microsystems~\cite{Haag}.  The
semigroup time evolution of the Gamow states is just an example of
this general time asymmetry.

\section*{Acknowledgments}

Augusto Garcia first drew our attention to the problem of mass
definition for resonances as discussed in \cite{Stuart} and
\cite{Bernicha.etal2}.  We are grateful to the following persons for
their contributions: A.~Mondragon for valuable
discussions on the subject, T.~Riemann for correspondence and
discussions which provided further phenomenological background on the
$Z$-boson, and D.~Dicus and G.~Lopez Castro for discussions and advice
on the
perturbation theory 
definitions of $Z$-boson mass and width and the problems with gauge
invariance and for reading and improving on the manuscript.  We
gratefully acknowledge the valuable support of the Welch Foundation.

\section*{Appendix: Time Asymmetric Quantum Theory of Scattering}

In this section we state the new hypothesis by which time asymmetric
quantum theory (TAQT) differs from the assumpitons of standard scattering
theory.  The dynamical (differential) equations, the algebras of
observables and the physical interpretation (probability) remain the
same as in time symmetric quantum theory in the Hilbert space \HS.
The only difference is that 
TAQT uses time asymmetric boundary conditions, and even these are
already implicit in the heuristic Lippmann-Schwinger (integral)
equations used for the calculation of the transition matrix
($T$-matrix).  Of the two
alternate ways of calculating the $T$-matrix~\cite{Newton}, we admit
as physically valid only the one which agrees with our intuitive
notion of causality.  In order to give this a mathematical
formulation, we need two Rigged Hilbert Spaces (RHS) with the same
Hilbert space \HS.

The relativistic Gamow vector is a precisely defined mathematical
object in the RHS formulation of quantum
theory.  This new quantum theory is not a drastic departure from the
usual Hilbert space (HS) formulation but an extension so that the 
theory includes Dirac kets \kt{E}\ and the solutions of the
Lippmann-Schwinger equation \mkt{E}\ and \pkt{E}\ which fulfill
outgoing and incoming boundary conditions, respectively.  The RHS and
the HS theory have the same (time symmetric) dynamical equations but
different boundary conditions.  The standard HS boundary conditions
are time symmetric: the space of in-states $\{\phi^+\}$ and the space
of out-states $\{\psi^-\}$ of scattering theory are identified with the
Hilbert space $\HS=\{\phi^+\}=\{\psi^-\}$ or at least with the same
subspace thereof $\{\phi^+\}=\{\psi^-\}\subset\HS$.

The RHS theory \emph{distinguishes} meticulously between prepared
\emph{states}
(in-states) $\{\phi^+\}$ and \emph{observables} (out-states) $\{\psi^-\}$ for
which it uses two RHS's 
of Hardy class
\begin{subequations}\label{eq:bc}
\begin{eqnarray}
\phi^+ & \in & \Phi_- \subset {\cal H} \subset \Phi_-^{\times} \\
\psi^- & \in & \Phi_+ \subset {\cal H} \subset \Phi_+^{\times}.
\end{eqnarray}
\end{subequations}
The space $\Phi_-$ ($\Phi_+$) is of Hardy class type in the
lower (upper) half plane; this means mathematically that the energy
wave functions $\pbk{E}{\phi}$ ($\mbk{E}{\psi}$) are well-behaved
Hardy class functions in the lower (upper) half plane,
$\mathcal{S}\cap\HS_-^2|_{\mathbb{R}_+}$
($\mathcal{S}\cap\HS_+^2|_{\mathbb{R}_+}$):
\begin{subequations}\label{innout}
\begin{eqnarray}
\phi^+\in\Phi_- &\Leftrightarrow&
\pbk{E}{\phi}\in\mathcal{S}\cap\HS^2_-|_{\mathbb{R}_+}\\
\psi^-\in\Phi_+ &\Leftrightarrow&
\mbk{E}{\psi}\in\mathcal{S}\cap\HS^2_+|_{\mathbb{R}_+}\label{caus}.
\end{eqnarray}
\end{subequations}
The notation $|_{\mathbb{R}_+}$ means the restriction to the positive
real line, i.e.\ the physical values of energy, and $\mathcal{S}$
denotes the Schwartz space.  The vectors $\phi^+$ represent states
that are prepared by the accelerator and the vectors $\psi^-$
represent observables or out-states that are defined by the detector.  The
dual spaces in the RHS's contain, in addition to the Dirac-Lippmann-Schwinger
kets $|E^{\pm} \rangle \in \Phi_{\mp}^{\times}$,
also Gamow kets $|E_R \pm i\frac{\Gamma}{2} ^{\pm} \rangle \in
\Phi_{\mp}^{\times}$.

This mathematical assumption of distinct RHS's of Hardy class for
states and observables is the new hypothesis from which 
the results that differ from the conventional theory follow.  At least
in the non-relativistic theory \cite{Harshman,Bohm2} one can
connect this mathematical assumption to the causality condition that a
state must be prepared at a time $t_0$ before an observable can be
measured at time $t>t_0$, and from this condition the Gamow vectors
can be naturally derived from the $S$-matrix.

\end{document}